# Describing the diverse geometries of gold from nanoclusters to bulk – a first-principles based hybrid bond order potential

Badri Narayanan,[1] Alper Kinaci,[1] Fatih G. Sen,[1] Michael J. Davis,[2] Stephen K. Gray,[1] Maria K. Y. Chan,[1*] and Subramanian K. R. S. Sankaranarayanan[1*]

[1] Center for Nanoscale Materials, Argonne National Laboratory, Lemont IL 60439 USA

[2] Chemical Sciences and Engineering Division, Argonne National Laboratory, Lemont IL 60439 USA

**ABSTRACT**

Molecular dynamics simulations using empirical force fields (EFFs) are crucial for gaining fundamental insights into atomic structure and long timescale dynamics of Au nanoclusters with far-reaching applications in energy and devices. This approach is thwarted by the failure of currently available EFFs in describing the size-dependent dimensionality and diverse geometries exhibited by Au clusters (e.g., planar structures, hollow cages, tubes, pyramids, space-filled structures). Owing to their ability to account for bond directionality, bond-order based EFFs, such as the Tersoff-type Bond Order Potential (BOP), are well suited for such a description. Nevertheless, the predictive power of existing BOP parameters is severely limited in the nanometer length scale owing to the predominance of bulk Au properties used to train them. Here, we mitigate this issue by introducing a new hybrid bond order potential (HyBOP), which account for (a) short-range interactions *via* Tersoff-type BOP terms and (b) long-range effects by a scaled Lennard-Jones term whose contribution depends on the local atomic density. We optimized the independent parameters for our HyBOP using a global optimization scheme driven by genetic algorithms. Moreover, to ensure good transferability of these parameters across different length scales, we used an extensive training dataset encompasses structural and energetic properties of a thousand 13-atom Au clusters, surface energies, as well as bulk polymorphs, obtained from density functional theory (DFT) calculations. Our newly developed HyBOP has been found to accurately describe (a) global minimum energy configurations at different clusters sizes as well as order of stability of various cluster configurations at any size, (b) critical size of transition from planar to globular clusters, (c) evolution of structural motifs with cluster size, and (c) thermodynamics, structure, elastic properties, and energetic ordering of bulk condensed phases as well as surfaces, in excellent agreement with DFT calculations and spectroscopic experiments. This makes our newly developed HyBOP a valuable, computationally robust but inexpensive, tool to investigate a wide range of materials phenomena occurring in Au at the atomistic level.

## 1. INTRODUCTION

Gold nanoclusters, owing to their exceptional chemical, optical, and electronic properties, are both fundamentally interesting and relevant to a wide range of applications, such as optoelectronics, bio-recognition, and catalysis.[1-5] There have, thus, been many theoretical and experimental investigations on neutral as well as ionic $Au_n$ clusters containing < 100 atoms.[2, 6-12] In particular, sub-nanometer gold clusters have been reported to exhibit pronounced catalytic activity (e.g., in low temperature oxidation of CO),[3, 13-15] and remarkable luminescence.[4, 16] Such exotic properties of $Au_n$ clusters are, at least in part, related to their geometric structure. Identification of the atomic configuration of these clusters has deservedly attracted a lot of attention in the recent past. A common strategy to extract structural information of gas phase $Au_n$ clusters is to employ a combination of spectroscopic measurements and theoretical predictions based on density functional theory (DFT) or tight-binding approaches.[2, 6-12] A surprising finding is the persistence of energetically favorable, planar $Au_n$ structures for configurations for up to $n$ ~12-14 atoms.[6, 9-12] Beyond $n$ = 12-14, three-dimensional structures are preferable. This is in contrast to alkali metal clusters, and other noble metal clusters, which remain planar only up to $n$ ~5-7 atoms.[17, 18] This peculiar behavior of gold has been attributed to relativistic effects, which enhance the hybridization of $5d$-$6s$ orbitals, and lead to overlap of $5d$ orbitals of neighboring Au atoms.[19, 20]

In addition to the existence of stable planar $Au_n$ clusters up to unusually large sizes, these clusters exhibit another fascinating structural phenomenon. They have been found to exhibit a size-dependent evolution of structural motifs from planar (up to $n$ ~12-14 atoms) to hollow cages ($n$ = 14 -18) to tubes (e.g. $n$ = 24, 32, 42 etc.) and finally to bulk-like compact structures (e.g., tetrahedra, icosahedra, dodecahera, etc.).[2, 8, 9, 21] This exotic behavior results in a wide diversity in the stable configurations for $Au_n$ clusters reminiscent of carbonaceous materials at the nanometer length scale. In contrast to the planar-to-globular transition, the occurrence of the globular motifs (i.e., cages, tubes) does not display a sharp transition.[21] For instance, symmetric $Au_n$ tubes can form at various sizes in the range $n$ = 24–72 with compact structures being more stable than tubes at intermittent sizes.[9, 21] Similarly, hollow cages can occur at sizes as high as $n = 50$.[22] From a practical standpoint, these structural motifs have been found to strongly influence cluster properties, such as catalytic activity.[3, 10, 14]

Despite significant strides in structural identification of $Au_n$ clusters, a fundamental understanding of the surface chemistry and dynamical processes governing formation of these clusters is still lacking. In addition, the global minimum energy configurations of the clusters in the mid-size regime ($n = 20 –100$) are not well understood. This can be primarily attributed to the intractability of exhaustive structural searches at these sizes in the framework of DFT even with the most efficient sampling methods (e.g., evolutionary algorithms, basin-hopping, etc.). Furthermore, since many isomers at a given cluster size (even at small sizes) are energetically close to each other (~20 meV/atom),[6] it is possible that they may undergo structural transitions under the influence of external stimuli, *e.g*., temperature fluctuations. The knowledge of such transitions and the associated mechanisms is still in its infancy. Global optimization of structures and molecular dynamics (MD) simulations based on empirical force fields (EFFs) or classical potentials provide an important route to address these issues.

The success of global optimization and MD techniques crucially depends on the ability of the employed EFFs to accurately describe interatomic interactions. The popular many-body potentials for Au available in the literature, e.g., embedded atom method (EAM)[23] or its variants like Sutton-Chen (SC)[24] and Gupta potentials,[25, 26] include a spherically symmetric effective electron density term in addition to pairwise interactions. This functional form accounts for metallic bonding, which relies on local electron density around each atom and consequently its coordination number. Such a framework works well for bulk phases of gold and other metals but fails for small clusters,[23-28] wherein bond directionality effects become important.[29]

Consequently, these spherically symmetric many-body potentials over-stabilize globular configurations of $Au_n$ even at sizes as low as $n = 6$.[28, 30, 31] The global energy-minimum structures predicted by these empirical force fields (EFFs) are either based on decahedral, icosahedral, and hexagonal prismatic motifs or are amorphous.[31-33] On the other hand, bond-order based EFFs (e.g, Tersoff-type BOP,[34] Reactive force field ReaxFF[35]) account for bond directionality via an angular dependence. However, the existing set of parameters have been fitted to a training set primarily consisting of thermodynamic, structural and elastic properties of bulk polymorphs of gold with limited data on clusters.[34, 35] These EFFs suffer from transferability issues when applied to $Au_n$ clusters, e.g., ReaxFF predicts the most stable isomer of $Au_8$ to be globular[35] in contrast to previous DFT calculations, which show that the most stable structure of $Au_8$ is planar.[6, 7] Therefore, these EFFs, at least with the available set of parameters, are not suitable for describing the diverse structural configurations exhibited by $Au_n$ clusters.

In this study, we systematically assess the performance of the available EFFs for $Au_n$ clusters (specifically around the planar-globular transition near $n = 13$) to identify the functional form that is best suited for describing diverse geometries. We find that the EFFs based on embedding electron density functions are inherently incapable of describing non-compact configurations (e.g., planar, hollow cages) due their treatment of many-body effects via spherically symmetric energy contributions and consequent neglect of bond directionality. Bond-order potentials, such as Tersoff-type BOP, account for these orientation effects via additional three-body interaction terms. Our assessment of EFF functional forms is consistent with previous reports, wherein bond directionality effects have been found to dominate the energetics/dynamics for small clusters.[29] Thus, bond-order formalism is essential for describing the various structural motifs exhibited by Au clusters. Among the bond-order potential forms available, the Tersoff-type BOP is computationally less intensive, and thereby, if re-parameterized, is best suited describe the size-dependent dimensionality effects and evolution of structural motifs in Au clusters. However, the long-range Au-Au interactions become more relevant in large clusters, surfaces, and bulk systems.[1] Since the Tersoff-type BOP accounts for short-range specifically nearest neighbor interactions alone (up to 3-body), an additional term in the potential energy function is necessary to capture long-range (LR) effects. To accurately capture size and dimensionality dependent LR interactions, we employ a pairwise LJ term that scales with the number of atoms within a prescribed radial distance from a given atom. We call this new potential model as Hybrid Bond Order Potential (HyBOP); it accounts for both short-range interactions *via* Tersoff-type BOP, and long-range effects using a scaled pairwise LJ.

Inspired by the nascent field of employing global optimization methods based on genetic algorithms (GA) for fitting force fields,[36-38] we use a combination of GA and local minimization methods (simplex[39, 40]) to optimize the values of the independent parameters in our HyBOP. For the parameterization, we employ a large training set comprising of DFT predicted energetic and structural properties of (a) $Au_{13}$ clusters, which represent structures with diverse local atomic coordination and amply sample the energy landscape, (b) bulk condensed phases of Au, including face-centered cubic (FCC) as well as other cubic polymorphs, and (c) energies of low-index surfaces of FCC gold. Our treatment of atomic interactions at both short as well as long-ranges enables accurate description of structure, energetics, and atomic scale dynamics across different length scales –from small clusters to surfaces and bulk condensed phases –making it suitable for wide range of Au based nano-scale systems. In particular, our newly developed HyBOP was found to accurately (a) reproduce the DFT predictions for the critical size at which planar-to-globular transition occurs, (b) predict global minimum energy structures at several sizes, (c) describe the size-dependent evolution of structural motifs in Au clusters, and (d) structure, thermodynamics, elastic properties, and energetic ordering of bulk Au polymorphs, in excellent agreement with DFT calculations as well as experiments. Furthermore, we employ this new EFF along with MD calculations to understand the dynamical processes at the atomic-scale that govern the agglomeration of a stable planar ($Au_7$) and a metastable globular (icosahedral $Au_{13}$) cluster to form an equilibrium configuration at the higher size, namely a $Au_{20}$ pyramid.

This paper is organized as follows: Section 2 describes the salient differences between the functional forms for EFFs using embedding functions and those based on bond-order formalism, the mathematical formulation of our HyBOP, DFT data used to construct the training set, and the techniques and strategies we employ for EFF parameterization. Section 3 reports our results on the performance of the EFFs available in the literature, the parameters of the HyBOP obtained in this study, the structural and energetic properties predicted by these parameters, and their success in describing the diversity of structural configurations displayed by $Au_n$ clusters. Section 4 provides a representative example for applying this newly developed EFF for studying agglomeration of clusters via long-time MD simulations, and highlights the successes/failures of this set of parameters. Finally, Sec. 5 summarizes the key findings and provides concluding remarks.

## 2. METHODS

*2.1 Functional form of the hybrid bond order potential (HyBOP)*

In the framework of HyBOP, the total potential energy is composed of partial energy contributions arising from atomic interactions that occur over both short- and long- distances, and can be written as:

$$V = V^{SR} + V^{LR},\qquad(1)$$

where $V_{SR}$ and $V_{LR}$ represents the energies associated with short-range (SR), and long-range (LR) interactions respectively. The SR interactions are treated using the Tersoff-Brenner formalism[34, 41-44] based on bond-order concept, in which $V^{SR}$ is expressed a sum of pairwise contributions:[44]

$$V^{SR} = \frac{1}{2}\sum_i \sum_{j\neq i} f_c(r_{ij})\left[f_R(r_{ij}) + b_{ij} f_A(r_{ij})\right],\qquad(2)$$

where $f_c(r_{ij})$, $f_R(r_{ij})$ and $f_A(r_{ij})$ are the cut-off function, repulsive and attractive interaction terms, respectively, between atoms $i$ and $j$ that are separated by a distance $r_{ij}$. The cut-off function limits the range of interaction to nearest neighbors for computational efficiency, and is given by:[44]

$$f_c(r) = \begin{cases} 1, & r < R - D \\ \frac{1}{2} - \frac{1}{2}\sin\left(\frac{\pi(r-R)}{2D}\right), & R - D \leq r < R + D \\ 0, & r \geq R + D \end{cases},\qquad(2)$$

where $R$ and $D$ are adjustable parameters. In this work, D is set to 0.2 Å following Ref. 34. The repulsive and attractive contributions to bond energy between a pair of atoms are defined to decay exponentially with separation distance:[44]

$$f_R(r) = A e^{-\lambda_1 r},\qquad(3)$$

$$f_A(r) = -B e^{-\lambda_2 r},\qquad(4)$$

where $A$, $B$, $\lambda_1$, and $\lambda_2$ are free parameters. The term $b_{ij}$ in Eq. 2 describes the bond-order around a pair of atoms $i$-$j$, which is dependent on three-body interactions given by Eqs. 5-7:[44]

$$b_{ij} = \left(1 + \beta^n \zeta_{ij}^n\right)^{\frac{-1}{2n}},\qquad(5)$$

$$\zeta_{ij} = \sum_{k\neq i,j} f_c(r_{ik}) g_{ik}(\theta_{ijk}) e^{\left[\lambda_3(r_{ij}-r_{ik})\right]},\qquad(6)$$

$$g(\theta) = \gamma \left( 1 + \frac{c^2}{d^2} - \frac{c^2}{d^2 + (\cos\theta + h)^2} \right). \tag{7}$$

In Eqs. 5–7, the parameters $\beta$, $n$, $\lambda_3$, $\gamma$, $c$, $d$, and $h$ are adjustable. The long-range interactions $V^{LR}$ are described using a scaled Lennard-Jones (LJ) function given by:

$$V^{LR} = \sum_i \sum_{j>i} 4\varepsilon_{ij} f_s(M_i) \left[ \left(\frac{\sigma_{ij}}{r_{ij}}\right)^{12} - \left(\frac{\sigma_{ij}}{r_{ij}}\right)^{6} \right], \tag{8}$$

where $\varepsilon_{ij}$, and $\sigma_{ij}$ are LJ parameters for a pair of atoms $i$ and $j$ that are a distance $r_{ij}$ apart. $f_s(M_i)$ is a scaling function that captures the dependence of LR contribution from a given atomic pair $i$-$j$, on the number of atoms within a prescribed radial distance $R_c^{LR}$ (set at 14 Å) from the atom $i$:

$$f_s(M) = \frac{1}{2}\left[ erf\left(\frac{M}{\kappa_1} - \kappa_2\right) + 1 \right], \tag{9}$$

where $\kappa_1$ and $\kappa_2$ are scaling parameters. **Figure 1** shows the dependence of the scaling function $fs(M)$ on $M$ for the values of $\kappa_1$ and $\kappa_2$ obtained in this work. The value of $f_s(M)$ is negligible for small clusters, and hence their energetics is dictated by the Tersoff type BOP terms only (*i.e.*, $V^{SR}$). As we move from small sized clusters to larger clusters (N > 150) and surfaces, the value of $M$ (i.e., number of atoms within $R_c^{LR}$) for most atoms increases beyond critical value ~90 and results in significantly higher values for $f_s(M)$; this in turn, causes the contributions from LR interactions to become significant. For bulk polymorphs, $f_s(M)$ reaches its maximum value of 1.0. We have also implemented the scaled LJ term (Eq. 8,9) in the popular MD simulation package LAMMPS.[45] The HyBOP contains 16 independent parameters ($R$, $A$, $B$, $\lambda_1$, $\lambda_2$, $\beta$, $n$, $\lambda_3$, $\gamma$, $c$, $d$, $h$, $\varepsilon$, $\sigma$, $\kappa_1$ and $\kappa_2$) that we optimize using the procedure detailed in Sec. 2.2.

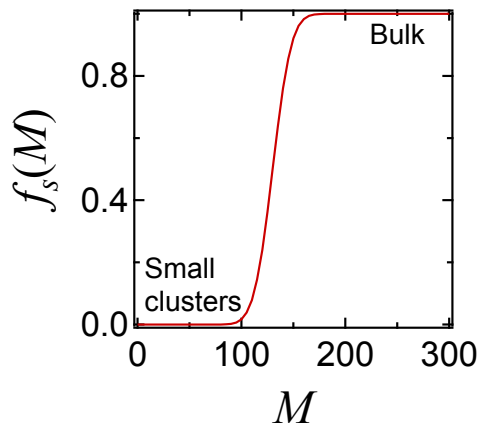

**Figure 1** Dependence of scaling function $f_s(M)$ on the number of atoms within a prescribed radial distance of a given atom. For small clusters $f_s(M) \approx 0.0$, while for bulk systems $f_s(M) = 1.0$.

*2.2 Parameterization strategy*

In development of EFFs, the training set for the fitting should constitute a good sampling of the possible structural configurations and their respective energies. For instance, to be able to predict both near equilibrium and far from equilibrium structures, a continuous range of energies from high to low should be included. In this study, we optimized all the 16 independent (free) parameters listed in Sec 2.1, using an extensive training set containing DFT computed (a) cohesive energies of a thousand $Au_{13}$ cluster configurations, and (b) equations of state of various cubic polymorphs of gold (*i.e*, face-centered, body-centered, simple, diamond, and $Cr_3Si$-type). The details of these DFT calculations are provided in Sec. 2.3.

To ensure adequate representation of the various possible coordination environments and cluster sizes, as well as the energy landscape, we employ two sampling techniques. First, we generate 1000 $Au_{13}$ nanoclusters by randomly placing atoms in a computational supercell such that all nearest-neighbor spacing are within 0.3 Å of the Au-Au bond length (2.88 Å) in the FCC gold crystal.[24, 35] These nearest neighbor distance constraints are imposed to avoid disjoint fragments (upper limit) and extremely large forces and accompanying convergence issues (lower limit). Among these 1000 random clusters, 100 are constrained to be planar while the remaining 900 are globular. We relax the atomic coordinates of these clusters using a conjugate gradient algorithm in the framework of DFT as described in Sec. 2.3. The relaxed configurations along with their energies are then employed in the training set. The DFT computed energies of these $Au_{13}$ configurations obtained *via* random sampling exhibit a Gaussian distribution centered at ~1.8 eV/atom as shown in **Figure 2(a)**; the most stable $Au_{13}$ cluster has a DFT energy ~2.1 eV/atom.[6] We find that the cluster configurations obtained by random sampling largely consist of fairly high-energy structures, and occur in a narrow phase space (the energies are within ~0.2 eV/atom of the most probable value). The near-equilibrium structures are particularly difficult to find using this sampling technique, even upon increasing the range of allowed closest Au-Au separations.

Next, to have better sampling of the atomistic system, we employ a structural search for low energy $Au_{13}$ clusters using evolutionary algorithms (GA sampling), wherein the cluster energies are evaluated using DFT. This technique complements the random sampling well, by identifying configurations in the potential energy landscape that are missed by random method. In particular, the energy distribution is nearly uniform, with energies spanning over a wide range ~1 eV [**Figure 2(b)**]. Importantly, near equilibrium structures are captured, which could not be obtained by random sampling. Overall, using these two sampling techniques, we obtained DFT computed cohesive energies for 1246 unique structures of $Au_{13}$; the energy difference between any two $Au_{13}$ configurations is atleast 8 meV/atom.

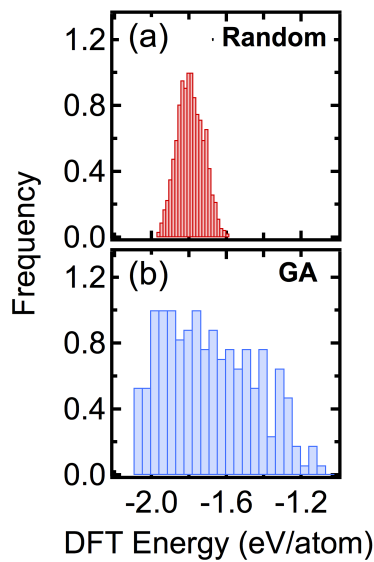

**Figure 2.** Energetic distribution of the $Au_{13}$ clusters obtained using (a) random, and (b) genetic algorithm (GA) sampling methods. A total of 1246 unique $Au_{13}$ cluster configurations were identified from these sampling techniques.

In addition to $Au_{13}$ clusters, our training set includes equations of state (*i.e*, cohesive energy *vs*. volume) for the various cubic polymorphs of Au computed using DFT. For these equations of state, when computing the energy as a function of volume, the lattice is subjected to ten different isotropic strains (*i.e*, equally along the three lattice vectors) with magnitudes in the range -10% (compressive) to 10% (tensile). It should be noted that although the PBE exchange correlation functional used in the present study for DFT calculations provides an accurate description of the structures and energies of clusters and surfaces, it underestimates the cohesive energy of bulk gold (DFT-PBE: -2.97 eV/atom;[46] Experiments: -3.81 eV/atom[47]). Despite this issue, DFT-PBE correctly identifies the energetic order of bulk phases.[35, 46] Thus, for fitting the parameters in the LR term, we employed the experimental value of cohesive energy of bulk FCC gold, and the DFT-PBE relative energies for other cubic polymorphs. Our DFT computed equation of states for these polymorphs were consistent with previous DFT studies.[46]

Using the training set described above, we optimize the BOP parameters by employing algorithms; the procedure is outlined in **Figure 3**. All the genetic operations in this procedure are performed by using the single/multi-objective genetic algorithm toolbox developed by Sastry and co-workers.[48] We begin the optimization procedure by generating a population of $N_p = 60$ parameter sets randomly, such that their values lie within physically allowable limits (for parameter search ranges refer Supplementary Information Table S1). Each set of BOP parameters (i.e., 12 adjustable ones) in this population is called a member. For each member $i$, we compute the BOP energies for all the structures in the training set using MD simulation package LAMMPS,[45] and evaluate the objective function $\Delta_i$ given by:

$$\Delta_i = \sum_j w_j \left( V_j^{BOP} - V_j^{DFT} \right)^2, \qquad (8)$$

where $V_j^{BOP}$ and $V_j^{DFT}$ are the BOP predicted, and DFT energies for the structure $j$ in the training set, while $w_j$ is the weighting factor. The weighting factors are prescribed such that the errors in the prediction of cluster energies control the optimization more than those for bulk. For clusters having DFT energies within 0.1 eV/atom of the most stable structure, $w$ is set to 1.0, those that lie between 0.1 to 0.3 eV are weighed at 0.75, while the higher energy clusters have $w = 0.5$. All the bulk structures are weighted with $w = 0.8$.

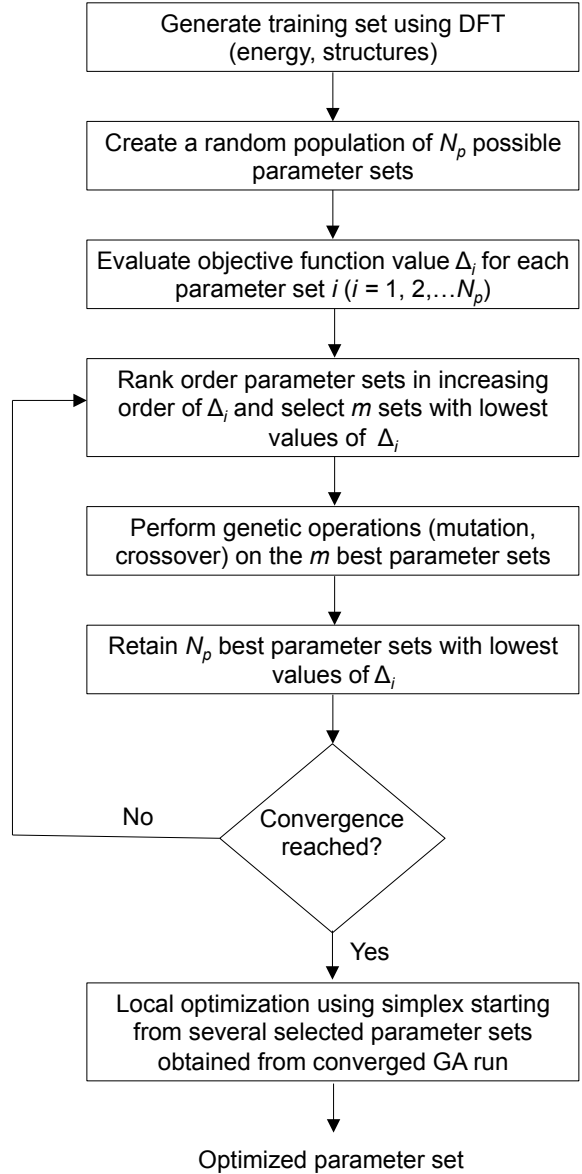

**Figure 3.** Flowchart describing the sequence of steps employed in this work for optimization of force field parameters

In a simpler sense, the objective function is the weighted sum of squares of errors of the BOP predictions. Members are then ranked in increasing order of $\Delta$. Among the $N_p$ members, the best $m = N_p/2 = 30$ members, i.e., the ones with lowest values of $\Delta$, are then subjected to genetic operations: crossover using the simulated binary method,[49, 50] or mutation via a polynomial of order 20.[51, 52] These mutations introduce sufficient diversity into the population, and avoid premature convergence of the GA run. The probability of a crossover operation is set to 0.9, while that of mutation is 0.1. After the genetic operations, both the old as well as new members are ranked by their Δ. The best $N_p$ parameter sets (members) were then chosen to constitute the next generation. Such an optimization routine ensures that only good parameter sets survive after each generation; upon repeating this workflow for sufficient generations, we sample diverse regions in the parameter space before the run converges, *i.e*, the values of Δ for at least 15 members in the population are below the prescribed tolerance. To ensure adequate sampling of the parameter space, we perform 10 separate GA runs starting with different random populations. From each of the converged GA runs, we choose 10 different parameter sets, whose errors in prediction are close or identical to that for the best set (*i.e*, one corresponding to the lowest Δ). To obtain the final parameter set, we run local optimization from these 10 sets using the simplex method[39, 40] until the difference between the Δ values for parameters in successive steps is less than $10^{-14}$.

*2.3 Details of the DFT calculations*

The DFT calculations are performed in the generalized gradient approximation (GGA) using the projector-augmented wave formalism as implemented in the Vienna Ab-initio Simulation Package (VASP).[53, 54] We employ Au PAW atom potential, without semi-core *p*-states, supplied with the VASP package. The exchange correlation is described by the Perdew-Burke-Ernzerhof (PBE) functional[55], which is known to describe the structure/energetics of Au clusters very well.[46] The plane wave energy cutoff is set to 230 eV for clusters and 500 eV for bulk polymorphs, while the Brillouin zone (BZ) is sampled by a Γ-centered Monkhorst-Pack grid. The plane wave energy cutoff of 230 eV was chosen for clusters based on convergence tests on a sub-set of configurations at various values of cutoff ranging from 200 to 520 eV; plane wave energy cutoff of 230 eV yielded cohesive energy values for Au clusters within 6 meV of those obtained at 520 eV (for details, see Supplementary Information). Furthermore, energy differences between structures are converged to better than 2 meV/atom at a planewave energy cutoff of 230 eV. For the planar clusters, a computational supercell of dimensions 30 Å × 30 Å × 20 Å is employed whereas for globular ones a 30 Å × 30 Å × 30 Å supercell is used. Such supercell dimensions ensured a vacuum of at least 10 Å along each direction to avoid spurious interactions across periodic boundaries. In all cluster calculations, the BZ is sampled by only the Γ-point. The atomic coordinates in the clusters are optimized partially (i.e., 18 ionic steps) using conjugate-gradient algorithm. Such a partial optimization is necessary for clusters far-away from equilibrium to ensure computational efficiency. For those near equilibrium, this procedure resulted in optimized geometries with force components on each atom < 0.01 eV/atom. For the bulk polymorphs, the computational supercell is composed of one conventional unit cell. A $k$-point grid of 16 × 16 × 16 is employed, which amounts to 165 $k$-points in the irreducible BZ. The atomic relaxations are performed using a conjugate gradient method until the force components on any atom are less than 0.01 eV/atom. We have also assessed the influence of including spin-orbit coupling (SOC), and van der Waals (vdW) dispersion effects using a vdW-DF functional[56, 57] on the PBE computed training data set. The inclusion of SOC, and vdW dispersion effects do not impact the energetic ordering of cluster configurations, global minimum configurations, Au-Au bond stretching, and Au-Au-Au angle bending energies. As such, inclusion of SOC does not alter relative stabilities of the clusters, which are of primary interest in force field fitting, consistent with earlier reports.[58] SOC inclusion only uniformly increases the cohesive energies by ~0.1 eV/atom while vdW increases it by ~0.03 eV/atom (for details, see Supporting Information). In addition, we note that a recent work on Au$_8$ clusters reports that DFT calculations within the PBE framework (without SOC/vdW corrections) reproduces the energetic ordering of Au cluster configurations derived from coupled-cluster calculations.[7] Moreover, since experimentally reported clusters are generally ionized, we checked for consistency between DFT evaluated energetic ordering for a set of $Au_{13}$ configurations in the presence of an extra $1e^-$ and that for neutral clusters; we performed these test on configurations from our DFT-GA searches. We found that any randomly chosen pair of $Au_{13}$ configurations displays a 98% probability of having the same energy ordering in the presence or absence of a net charge. This instills confidence in the accuracy of the energetic order of $Au_{13}$ configurations as well as the global minimum structure obtained from DFT-PBE.

## 3. RESULTS

*3.1 Performance of the existing FFs for Au nanoclusters*

The capability, transferability, and limitations of an EFF are, to a great extent, dictated by the functional form describing inter-atomic interactions. For instance, pairwise potentials, such as a Morse potential, cannot describe directional bonding in covalent materials owing to the absence of energy contributions from 3-body and higher order interactions.[43, 44] In terms of the present study, it is, therefore, crucial to identify the most important characteristics of a functional form that can describe the energetics/dynamics accurately over the all the length scales from small clusters up to surfaces and bulk Au. To accomplish this, we first assessed the predictive capability of the existing Au EFFs for few atom gold nanoclusters.

**Figure 4** compares the cohesive energies of 1246 $Au_{13}$ clusters [**Figure 2**] as predicted by SC,[24] EAM,[23] Tersoff-type BOP,[34] and ReaxFF[35] with their corresponding values obtained from DFT. All these EFFs overestimate the cohesive energies of $Au_{13}$ clusters [**Figure 4**], with mean absolute errors in energies (as compared to DFT), δ, ranging from 0.4–1 eV/atom with a standard deviation of 0.05–0.1 eV/atom [Figs. 3(a-d)]. Furthermore, the values of δ are higher for globular clusters (~0.5—1.2 eV/atom) than the planar ones (~0.2—0.8 eV/atom). More importantly, we note that the bond-order potentials (*i.e.*, ReaxFF and BOP) describe the energetics of Au clusters better than those that account for many-body effects via an embedded electron density function (i.e., SC and EAM).

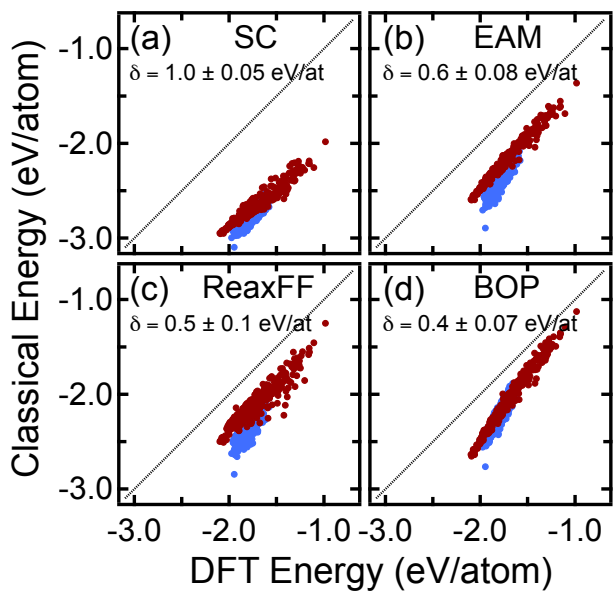

**Figure 4.** Comparison of the cohesive energies of 345 planar (red), and 901 globular (blue) $Au_{13}$ nanoclusters computed using DFT (PBE-GGA) against those predicted by various FFs available in the literature (a) Sutton-Chen [Ref. 24], (b) Embedded atom method [Ref. 23], (c) Reactive Force Field [Ref. 35], and (d) Tersoff-type BOP [Ref. 34]. The mean absolute errors (δ) in the predicted energies as compared to the DFT values, along with the standard deviations, are also provided in each panel.

In addition to cohesive energies, it is essential to assess the performance of these EFFs in reproducing the energetic ordering of structural isomers at a given cluster size. **Figure 5** compares the energetic ordering of 5 isomers of $Au_{13}$ clusters predicted by existing EFFs with those known by DFT calculations.[6, 10] The $Au_{13}$ isomers displayed in the bottom panel of **Figure 5** have been previously identified, using DFT calculations, as configurations that lie near the global energy-minimum.[6] Of these, the isomers named $P_i$ ($i = 1–4$) are planar; here the subscript indicates the DFT order of stability (1: most stable; 4: least stable), while $I_h$ is globular with icosahedral symmetry.

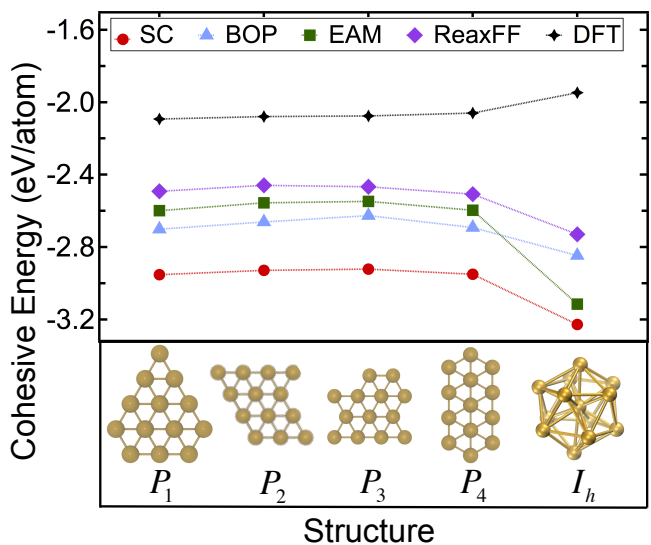

**Figure 5.** Predictive power of the available empirical EFFs for $Au_{13}$ configurations that are close to the DFT-predicted global energy minimum structure ($P_1$). The cohesive energies of four planar structures ($P_1$, $P_2$, $P_3$ and $P_4$) and one icosahedron ($I_h$) are computed with EFFs and compared with those obtained by DFT. All the available EFFs predict the globular $I_h$ to be the most stable structure for $Au_{13}$ in contrast to DFT (which predicts planar $P_1$ to be the global energy minimum).

**Figure 5** shows that all the current EFFs predict $I_h$ to be the most stable structure for $Au_{13}$. This is in contradiction to the DFT predicted global energy-minimum configuration, which is planar $P_1$.[6] Even among the planar isomers, the existing EFFs do not preserve the DFT order of stability (e.g., SC predicts $P_4$ to be most stable planar structure). Evidently, currently available classical

FFs fail to capture the dimensionality effects at the nanoscale; these deficiencies could possibly be related to either (a) inherent limitations of their functional form, and/or (b) the lack of sufficient data on clusters in the training set used to parameterize them.

We also assess the dynamic stability of planar $Au_{13}$ clusters in the framework of the currently available EFFs *via* long time MD simulations (using LAMMPS[45]). In these simulations, the planar clusters ($P_1 - P_4$) are individually heated from 0 K to 300 K over 5 ns and subsequently held at 300 K for an additional 5 ns. We find that in the framework of the EFFs which incorporate multi-body effects *via* embedded density functions, *i.e.*, SC and EAM, even a slight perturbation (as low as ~2 K) to the plane of $Au_{13}$ causes it to collapse into a centered icosahedron. In other words, these EFFs predict the planar forms of $Au_{13}$ to be dynamically unstable. We note that such a collapse does not occur during a local structural optimization (e.g., *via* conjugate gradient algorithm). This indicates that the planar forms of $Au_{13}$ lie in either a very shallow minimum or at saddle points in the energy landscape of the SC and EAM potentials. In contrast, our canonical *ab initio* MD simulations (0.01 ns long) show that the planar forms remain structurally stable even at higher temperatures ~500 K.

Using global optimization procedure outlined in Sec. 2.2, we re-parameterized the SC potential to provide better description of Au nanoclusters (for details, see Supplementary Information). From these parameterization attempts, we find that EAM type potentials cannot describe the structural diversity exhibited by few-atom Au clusters, while still providing reasonable predictions of the bulk properties. This is consistent with previous reports wherein SC parameters were optimized to reproduce DFT computed Au-Au bond stretching, and Au-Au-Au angle bending energies;[59, 60] although, the parameters developed in these works could describe dimers and trimers well, they do not perform well for clusters containing > 8 atoms.[60]

Furthermore, no significant improvement in the dynamic stability of planar clusters could be achieved even upon re-parameterizing the SC potential using our cluster training set. Thus, the framework of EAM type EFFs, which employ a spherically-symmetric effective electron density to account for metallic bonding, is not suitable for describing the diverse structural configurations of Au nanoclusters.

On the other hand, existing bond-order based potentials (BOP [Ref 34] and ReaxFF [Ref 35]) capture bond-directionality via incorporating angular (BOP) and/or dihedral (ReaxFF) dependences. Such a framework enables these EFFs to preserve the planarity of $Au_{13}$ clusters for a long time (~4 ns) at temperatures up to 300 K. This indicates that the planar configurations lie in well-defined minima in the potential energy surface of these potentials. At 300 K, the planar clusters eventually transform into 3D globules ($I_h$ using ReaxFF; disordered using BOP), which is not very surprising since these potentials were primarily parameterized to bulk properties. The barrier associated with collapse of planar Au clusters, and the relative energetic ordering between planar and globular forms is not well described with the current set of BOP/ReaxFF parameters; this is related to the lack of sufficient information about Au clusters in their training set. Nevertheless, the dynamic stability of the planar clusters in the framework of the bond-order based EFFs is much better than that with EAM or SC potentials.

Evidently, empirical potential functional forms based on bond order formalism are essential to describe the energetics/atomic-scale dynamics and structural diversity in small Au clusters. Among the bond-order potentials available in literature, the Tersoff-type BOP holds distinct advantage over ReaxFF owing to its simpler mathematical formulation, and consequently, lower computational costs.[35, 44] Although bond orientation effects are dominant in clusters (which can be accurately described by Tersoff-type BOP), previous studies report that long-range Au-Au interactions are significantly high in large clusters, surfaces and bulk polymorphs. Hence, to enable accurate treatment of Au across all length scales and dimensionality (*i.e.*, clusters, surfaces and bulk) we introduce a HyBOP that include bond-order terms as well as pairwise LJ terms to describe Au-Au interactions at short- and long-ranges respectively [Sec. 2.1].

*3.2 Parameterization of hybrid bond order potential*

An earlier work (Ref. 34) reports BOP parameters for Au; these parameters were obtained by fitting to a training set containing cohesive energies, structures, elastic constants, and surface energies for bulk polymorphs of Au derived from DFT calculations with local-density approximation and experimental measurements. Although this set of parameters describes the structure, thermodynamics, and elastic properties of bulk Au accurately, it lacks transferability in the nano-scale size regime. Consequently, these parameters do not capture the energetics, size-dependent changes in structural motifs, and order of stability of various structural configurations at a given size for Au nanoclusters [**Figure 4**, **Figure 5**]. Therefore, we parameterized the BOP parameters for SR interactions from scratch by employing a large training data set, which adequately samples the various structural configurations in Au nanoclusters [Sec. 2].

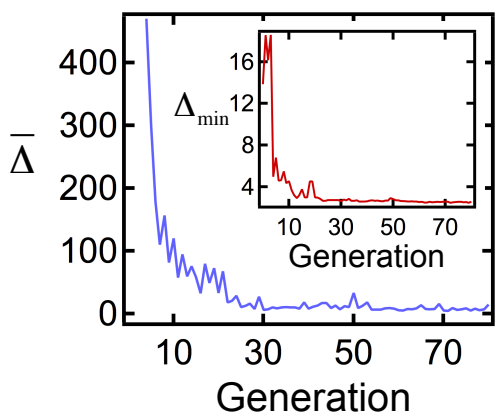

**Figure 6.** Evolution of the average and minimum value (inset) of the objective function during global optimization of BOP parameters for Au nanoclusters.

Using GA optimization algorithms [**Figure 3**], we obtain the independent parameters for the HyBOP potential by training against cohesive energies of 1246 representative clusters, as well as equation of state of cubic polymorphs of gold evaluated by DFT calculations [Sec. 2]. **Figure 6** shows the evolution of the mean ($\bar{\Delta}$) and minimum ($\Delta_{min}$) values of the objective function in a typical run of our global optimization of BOP parameters. These runs converge within ~30 generations, which involves ~2.5×10$^6$ BOP energy evaluations (*i.e*, for the structures in the training set). The stationary values of $\bar{\Delta}$ and $\Delta_{min}$ at the end of a typical optimization run are ~7 and ~2 respectively. Physically, this means that the best parameter sets (that result in an objective value of $\Delta_{min}$) can predict energies of Au nanoclusters to within ~50 meV/atom of the DFT values.

**Table 1.** Hybrid bond order potential parameters obtained in this study

| Parameter | Value |
|---|---|
| *Short range interactions* | |
| $\gamma$ | 1.053×10$^{-3}$ |
| $\lambda_3$ (Å$^{-1}$) | 2.0522 |
| $c$ | 3.3592 |
| $d$ | 0.1647 |
| $h$ | -0.9942 |
| $\beta$ | 0.99994 |
| $n$ | 0.99992 |
| $\lambda_2$ (Å$^{-1}$) | 1.6069 |
| $B$ (eV) | 247.4785 |
| $R$ (Å) | 3.1994 |
| $\lambda_1$ (Å$^{-1}$) | 3.1335 |
| $A$ (eV) | 5453.12895 |
| *Long range interactions* | |
| $\varepsilon$ (eV) | 0.1509 |
| $\sigma$ (Å) | 2.6731 |
| $\kappa_1$ | 20 |
| $\kappa_2$ | 6.5 |

Interestingly, we find that once our optimization routine converges, several dissimilar parameter sets appear, which give objective values close (even identical) to the lowest Δ. Although these parameter sets result in similar overall objectives (i.e., sum of squares of errors in predictions), their performance can be vastly different for structure and energy predictions for clusters. For instance, one of the parameter sets from our converged global optimization could predict cluster energies within 30 meV/atom of the DFT values; this set, however, does not correctly reproduce DFT predictions for global-energy minimum configuration of Au$_{13}$ cluster.

It is, therefore, essential to choose a parameter set from the post-convergence ones, which can describe the cohesive energies, thermodynamic ordering, and dynamics equally well. Furthermore, these parameter sets do not necessarily represent local minima in objective values in the multi-dimensional space of the independent parameters. To ensure that our final parameter set is indeed a local minimum, and to guarantee reasonable success in predictions for structure/energies of Au clusters and bulk phases, we performed local optimization (using the simplex algorithm[39, 40]) starting from several parameter sets that appear post-convergence in our genetic (global) optimization runs. After the local optimizations, the performance of the optimized parameter sets is evaluated on the following fronts: (a) energetic ordering of Au$_n$ ($n$ = 12-14) clusters, (these sizes, as aforementioned, are near the DFT predicted planar-globular transition size), (b) predictive power of cohesive energies of clusters, and (c) elastic stability, and thermodynamic ordering of the bulk cubic polymorphs. The set of parameters that perform the best on all these fronts are shown in **Table 1**.

*3.3 Performance of our newly developed HyBOP for small clusters*

Next, we assess the performance of our newly developed HyBOP [**Table 1**] in terms of describing the energetics of small Au clusters. For the Au$_{13}$ clusters employed in the training set, the energies predicted by our HyBOP parameters are in close agreement with the DFT values [**Figure 7 (a)**] with mean absolute errors of 0.07 eV (as opposed to 0.4 eV using the set of BOP parameters in Ref. 34). Note that the LR Au-Au interactions are negligible for such small sizes [**Figure 1**]. Thus, the improved performance of our HyBOP model for small cluster is due to the enhanced predictive power of our newly obtained BOP parameters for SR interactions. We also compare the ability of the two sets of BOP parameters in describing the energetic ordering of various structural Au$_{13}$ isomers. To accomplish this, we plot the energies of 10 planar and 10 globular Au$_{13}$ configurations that occur near the global energy minimum [**Figure 7(b)**]. These structures are obtained from our GA sampling using DFT[61] (See Supplementary Information for details of the technique). Evidently, our newly obtained BOP parameters for SR interactions describe the energetic ordering of the Au$_{13}$ isomers better than those in Ref. 34. In particular, our parameter set predicts the planar $P_1$ structure to be the global minimum configuration for Au$_{13}$ identical to earlier DFT calculations.[6, 10, 20] In contrast, the original set of parameters predicts the icosahedron $I_h$ to be the most stable structure. Furthermore, we find that even for Au$_n$ clusters of sizes other than $n = 13$ (which are not included in the fitting process), the energies predicted by our BOP parameters are within ~0.15 eV/atom of the DFT values.

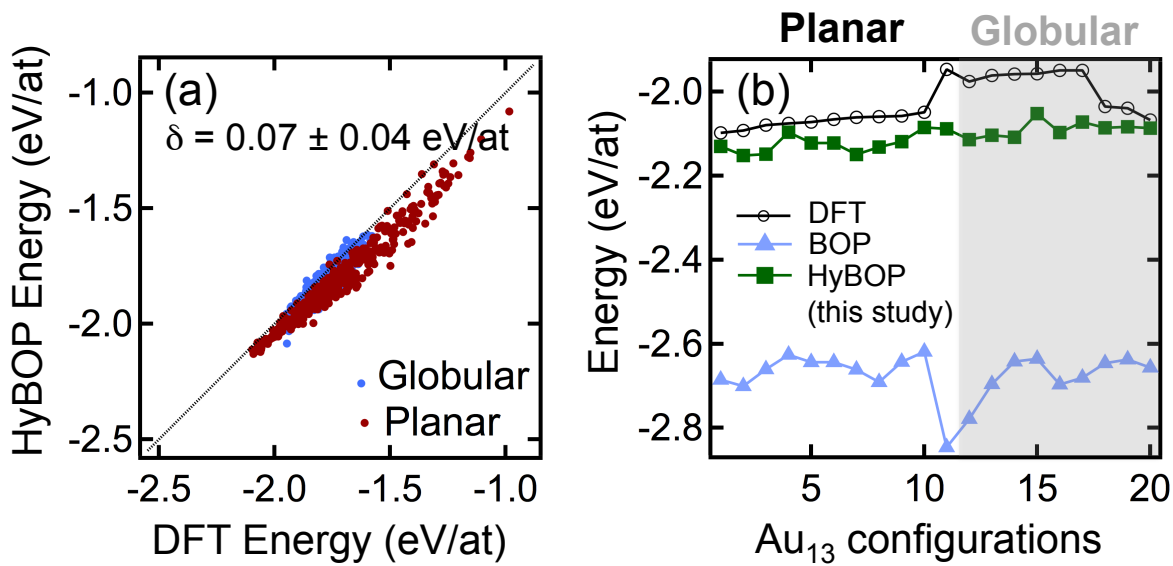

**Figure 7.** Performance of the BOP parameters obtained in this study for Au nanoclusters. (a) Comparison of the cohesive energies predicted by the re-parameterized BOP for 1246 planar (red) and globular (blue) $Au_{13}$ configurations against those computed by DFT; (b) Cohesive energy of 20 near global minimum $Au_{13}$ configurations evaluated using DFT and BOP with parameters fitted to Au bulk Ref. 34, and those obtained in this study [**Table 1**]. The configurations plotted in the shaded region are globular.

Although the newly developed HyBOP is significantly better at reproducing DFT data than the current EFFs, it does not capture all of the energy trends for the near minimum $Au_{13}$ configurations [**Figure 7**]. This is because the energy difference between the configuration pairs whose relative stability are not correctly captured (with respect to DFT standard) are ~30 meV/atom, which is lower than or similar to the errors (deviations from DFT value) associated with predictions of HyBOP [~40 meV/atom]. Since one does not expect DFT-PBE to exceed (or even reach) chemical accuracy (~40 meV/atom), the EFF is performing essentially at the limit of DFT accuracies. Moreover, the newly developed HyBOP reproduces the DFT order of stability for a randomly chosen pair with a probability of ~90%. Thus, our newly parameterized BOP is capable of describing the energetics of Au nanoclusters more accurately than existing EFFs.

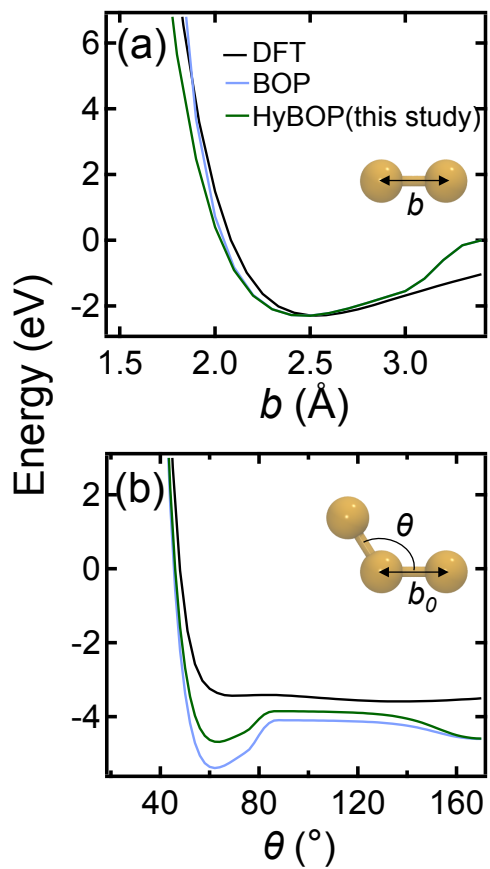

**Figure 8.** Comparison of the HyBOP predicted energies for (a) Au-Au dimers as a function of separation b, and (b) Au-Au-Au trimers as a function of the valence angle with DFT calculations. For the calculations shown in panel (b), the closest Au atoms are held at the equilibrium bond distance $b_0$ at the corresponding level of theory.

To understand why our newly parameterized BOP (and, in turn, our HyBOP) fares much better than the original Tersoff-like BOP[34] for small clusters, we compare the predicted Au-Au bond stretching as well as Au-Au-Au angle bending energies from these two sets with those obtained from DFT [**Figure 8**]. The BOP predicted energies (both new and original parameters) of Au dimer as a function of the Au-Au separation distance *b* are in good accordance with DFT calculations, as shown in **Figure 8(a)**. Both sets of BOP parameters are found to predict identical values for equilibrium Au-Au bond length (2.46 Å) in excellent agreement with DFT calculations (2.53 Å) as well as previous experiments (2.47 Å). A slight distinction, however, appears between the two BOP parameters at shorter distances. The original BOP employs Ziegler-Biersack-Littmark universal repulsive potential at small separations,[62] which results in a hard repulsion. To circumvent this, we use the pairwise repulsive energy contribution intrinsic to the BOP potential form [**Eq. 3**]. This leads to a softer repulsion but our new set of parameters ensure close reproduction of the DFT evaluated dimer energies even at small separations [**Figure 8(a)**].

The Au-Au-Au trimer energies as a function of the valence angle $\theta$ are plotted in **Figure 8(b)**. At the level of DFT, beyond the repulsive region at low values of $\theta$, the angle bending energies vs $\theta$ remains largely flat after $\theta \sim 60°$ within an additional shallow minima at ~150°. Such a flat energy landscape is extremely difficult to reproduce with EFFs while still preserving the ability to describe the structure/dynamics in few atom clusters and bulk with reasonable accuracy. As a consequence, the BOPs (both new and original parameters) reproduce the overall DFT trends for angle bending energies, with a well-defined minima at $\theta \sim 64°$, and a shallow one at ~160°. Our new set of parameters predict a much smaller well depth at $\theta \sim 64°$ as compared to those in Ref. 34 [**Figure 8(b)**]. Consequently, with our new parameter set, the two minima (at 64°, 160°) possess similar energies; this contributes greatly to the ability of the HyBOP parameters developed in this work to accurately describe the various structural motifs exhibited by Au clusters.

## 4. DISCUSSION

*4.1 Global minimum energy (GM) configuration based on our BOP for small nanoclusters*

In addition to describing the energetics well, another crucial yet challenging aspect of an EFF is its ability to accurately capture the global minimum energy (GM) configuration for nanoclusters at different sizes. This is a particularly difficult problem for cluster sizes that have not been represented in the training set employed to parameterize the EFF. Consequently, in most cases, capability of an EFF in predicting GM structures for nanoclusters at various sizes can be regarded as a true test of its transferability. Indeed, our newly developed HyBOP shows good success in predicting GM configurations for nanoclusters at several sizes. **Figure 9** shows GM structures of Au$_n$ clusters at selected sizes predicted by the HyBOP parameterized in this work using genetic algorithm based structural searches [for details of the technique please refer Supplementary Information]. From these extensive searches, we found that the GM structures are planar for Au$_n$ clusters containing $n < 14$ atoms; at $n = 14$ and beyond, globular GM structures appear. This predicted critical size at which the transition from planar to globular configurations occurs (14 atoms) is identical to previous DFT calculations,[6, 8, 9] as well as ion mobility and spectroscopy measurements.[8, 11] Aside from the correct prediction of this critical size, we find that the GM structures predicted by our HyBOP are also in excellent accordance with previous DFT and experimental reports. For instance, the HyBOP predicted planar GM structures for sizes up to 13 atoms are identical to the ones reported in Ref. 34 using DFT calculations. More importantly, among the globular clusters, our new parameters can successfully reproduce the evolution of various structural motifs with cluster size as observed in DFT calculations and spectroscopic experiments.[2, 6-12] In the size range $n = 14-17$, HyBOP predicts hollow cages [e.g., Au$_{14}$ in **Figure 9**] as GM configurations, which are analogous to previous DFT based basin-hopping searches and photoelectron spectroscopy.[8] At $n = 20$, the HyBOP developed in this study indicates a pyramid like space-filled structure [**Figure 9**] to be the most stable, similar to previous spectroscopy measurements.[2]

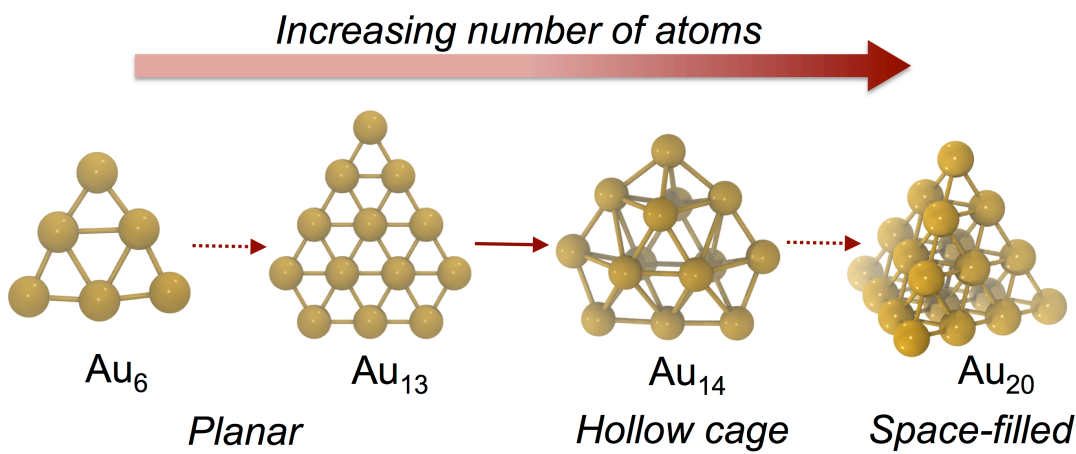

**Figure 9.** Global energy minimum configurations of Au nanoclusters at different sizes as predicted by our newly parameterized bond order potential.

Interestingly, our HyBOP based GM structural searches in the size range $n = 12 - 14$ indicate the existence of several planar as well as globular isomers, which are energetically within ~30 meV/atom of GM. This is consistent with previous DFT investigations and can be held responsible for the debate surrounding the critical cluster size at which planar-globular transition occurs (reported values range from 12—14). Furthermore, this indicates that our HyBOP parameters are well equipped to capture the transition/intermediate states, which arise during various dynamical phenomena in Au nanoclusters. A few examples of such phenomena include formation of Au clusters either atom-by-atom or *via* combination of smaller clusters, phase transition between different isomers at a given cluster size, and melting. From the parameterization point of view, it is particularly interesting to note that our HyBOP parameters obtained by training against thousand Au clusters of one size, namely 13 atoms (alongside the equation of state of bulk Au polymorphs) could accurately capture the evolution of structural motifs in Au nanoclusters with increasing number of atoms. The structure/energy information at other cluster sizes was not required.

*4.2 Dynamics of coalescence of planar and space-filled nanoclusters*

To demonstrate the ability of our HyBOP parameters to capture atomic-scale dynamic processes, we employ long-time MD simulations to study the formation of compact $Au_{20}$ pyramid via coalescence of two clusters: (a) space-filled $Au_{13}$ icosahedron and (b) planar $Au_7$ [**Figure 10**]. As mentioned earlier, the icosahedral structure is not the GM for $Au_{13}$ [Sec. 3.3]. It is still a low-lying isomer (~40 meV/atom higher than the GM), which can co-exist with the planar $Au_{13}$ in typical experiments. The planar configuration of $Au_7$, on the other hand, is the GM configuration. Initially, the $Au_{13}$ and $Au_7$ clusters are placed sufficiently close to each other in vacuum, such that the nearest atoms of the two clusters are ~3.5 Å apart [**Figure 10(a)**]. The initial velocities for the atoms were chosen randomly from a Maxwell-Boltzmann distribution consistent with ambient temperature (300 K) and there is zero overall momentum.

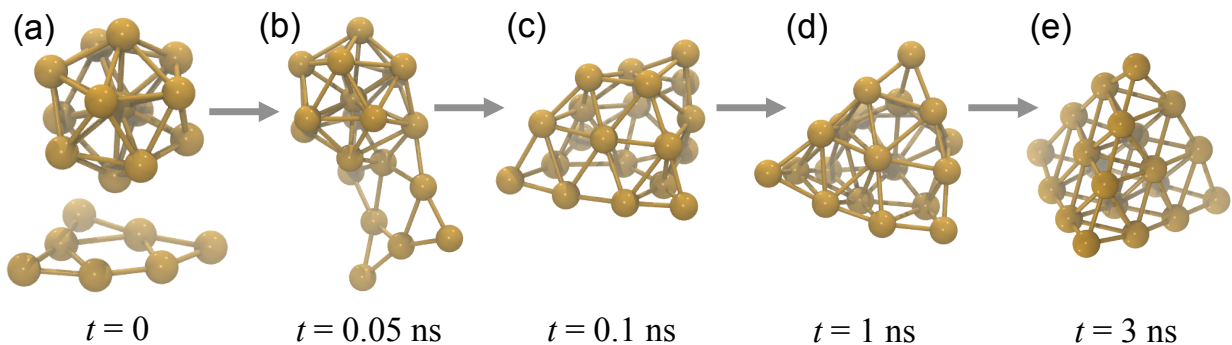

**Figure 10.** Dynamics of coalescence of an $Au_{13}$ icosahedron and a planar $Au_7$ at 300 K as described by the BOP parameters developed in this study.

We monitor the evolution of the atomic configuration in canonical MD simulations for 10 ns with a timestep of 1 fs; temperature is maintained at 300 K via Nosé-Hoover thermostat. During these simulations, we find that the clusters attach within ~0.04 ns, and the overall configuration looks like an icosahedron with a planar appendage, as shown in [**Figure 10(b)**]. This planar appendage causes the icosahedron to open up resulting in a distorted hollow cage like structure at ~0.1 ns [**Figure 10**(c)]. Once the icosahedral part of the structure is opened up, the collection of atoms sample through a wide range of hollow cage like structures [**Figure 10(c,d)**]. These hollow-cage like configurations were found to appear frequently in our HyBOP based GM structural search, since they are energetically close to the most stable compact pyramid (~20 meV/atom). Eventually, these hollow cages transform into a compact $Au_{20}$ pyramid at t ~3 ns, which is the GM configuration [**Figure 9**, **Figure 10(e)**]. At this point, the system reaches equilibrium; thereafter, for the remainder of the simulation (up to 10 ns) only thermal perturbations were observed in the pyramidal structure. It is important to note that such long timescales (3 ns) are currently not tractable by *ab initio* MD methods. This further justifies the need for an accurate empirical potential like our newly developed HyBOP for studying formation dynamics even for very small clusters (containing few tens of atoms).

*4.3 Transferability of HyBOP parameters for large clusters and surfaces*

To assess the accuracy and transferability of our newly obtained HyBOP parameters for large Au nanoclusters, we compare the cohesive energies of ten representative configurations of $Au_{40}$ and $Au_{75}$ clusters predicted by our HyBOP, and Tersoff-type BOP parameters (Ref. 34) with those derived from DFT calculations [Error! Reference source not found.]. The cluster configurations are chosen in such a way that they span a sufficiently wide range of cohesive energies (~0.4 eV/atom) at both the sizes. As indicated by Error! Reference source not found., our HyBOP predicts the cohesive energies of $Au_{40}$ and $Au_{75}$ clusters better than BOP (Ref. 34). For $Au_{40}$ clusters, we found that the average absolute error in predicted cohesive energies (from the DFT values) using our HyBOP is 0.08 eV/atom, and the standard deviation is 0.03 eV/atom, which is much lower than that for the original BOP (0.6±0.04 eV/atom). The errors in prediction associated with HyBOP and original BOP exhibit similar behavior for $Au_{75}$ clusters as well (this study: 0.04±0.02 eV/atom; BOP (Ref. 34): 0.6±0.04 eV/atom). However, both these potential models reasonably reproduce the DFT energetic ordering of the various cluster configurations at a given size. Each of these models predict the correct relative stability between a pair of randomly chosen cluster configurations (at a given size) with a probability of ~0.9 (for $Au_{40}$), and ~0.98 (for $Au_{75}$).

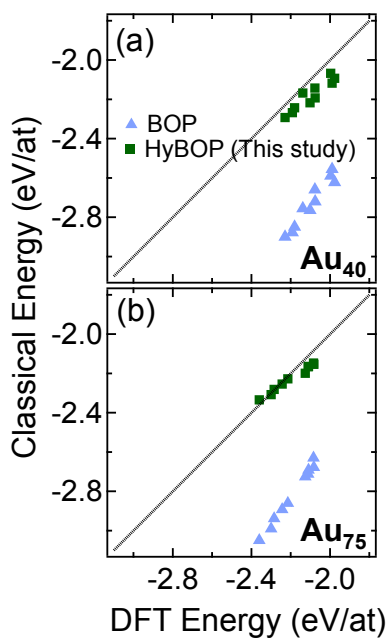

**Figure 11** Comparison of the cohesive energies of representative configurations of (a) $Au_{40}$, and (b) $Au_{75}$ clusters predicted by Tersoff-type BOP reported in Ref. 34 (blue triangles), and HyBOP developed in this study (green squares) with those obtained from DFT calculations.

The structure of large Au clusters depends mainly on surface faceting, and their equilibrium shape is composed of facets of low index planes such that the overall surface energy is minimized for a given cluster volume. Such a shape can be obtained by the well-known Wulff construction. Hence, the structure of large Au clusters is largely dictated by the relative energies of low index surfaces of FCC gold. **Table 2** compares the predicted surface energies for the three prominent low index surfaces of gold, namely 111, 100, and 110 using our newly developed HyBOP and existing EFFs with those derived from DFT calculations and experiments. Evidently, the surface energies predicted by our HyBOP are in better agreement with DFT calculations and experiments, as compared to popular force fields available for Au,[23, 34, 35, 63] including the original BOP.[34] Consequently, the surface faceting and shape of large nanoparticles as determined by Wulff construction using the HyBOP predicted surface energies is in excellent accordance with that obtained by DFT calculations [**Figure 12**]. It is interesting to note that the cluster shape derived from DFT is composed of 111 and 100 surfaces alone; in contrast, the Wulff shapes predicted by all the popular Au EFFs (except Tersoff-type BOP) possess significant 110 facets. Indeed, our HyBOP does not exhibit 110 facets in its predicted Wulff plot [**Figure 12**]. Moreover, the overall cluster shape predicted by our HyBOP is nearly identical to that obtained from DFT, and outperforms all the popular Au EFFs including the original BOP.

Owing to its bond order formalism, our HyBOP is well equipped to describe mixed metallic-covalent systems as shown by numerous examples in the literature.[64-66] Such heterogeneous systems are prevalent in Au nano-catalysis, e.g, synthesis of hydrocarbons via Fischer-Tropsch reaction catalyzed by Au nanoclusters.[67] However, to extend the capability of our HyBOP for Au clusters to such heterogeneous systems, it is essential to determine optimum parameters to describe the cross-interactions, e.g., Au-C, Au-H, and Au-O. The optimization strategy outlined in the present work (i.e, a combination of global optimization using genetic algorithms, and local optimization via Simplex algorithm) is applicable to parameterize these interactions; however, optimizing the HyBOP parameters for these cross interactions is beyond the scope of the current work.

**Table 2** Comparison of the surface energies of various low index faces of FCC gold predicted by our newly developed HyBOP potential with those calculated using other popular EFFs for gold. The corresponding values derived from DFT calculations and experimental measurements are also provided.

| Surface Energies (J/m$^2$) | HyBOP (This study) | EAM (Ref. 23) | SC (Ref. 63) | BOP (Ref. 68) | ReaxFF (Ref. 35) | DFT (This study) | Experiment |
|---|---|---|---|---|---|---|---|
| 111 | 1.76 | 0.79 | 0.58 | 0.67 | 1.76 | 1.39 | 1.5$^c$ |
| 100 | 2.04 | 0.92 | 0.66 | 0.90 | 1.99 | 1.62 | (average) |
| 110 | 2.16 | 0.98 | 0.70 | 0.97 | 2.10 | 1.75 | |

$^a$Ref. 47

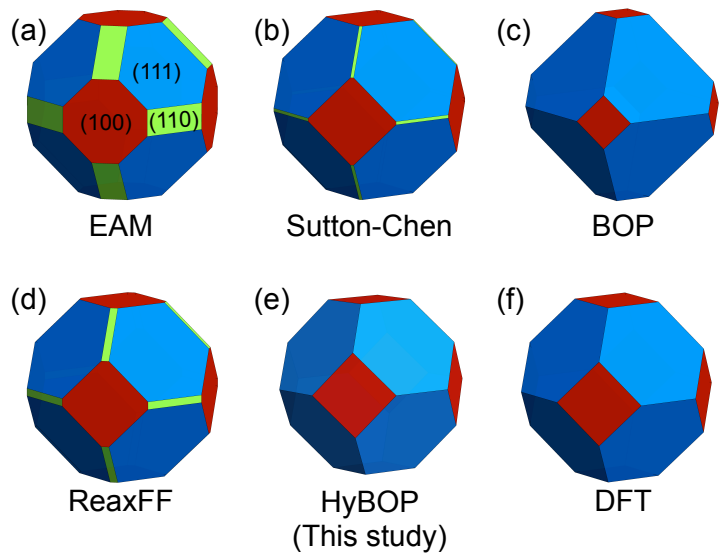

**Figure 12** Wulff construction of cluster shapes for Au using surface energies (provided in Table R2) predicted by (a) EAM [Ref. 23], (b) Sutton-Chen [Ref. 63], (c) BOP [Ref. 34], (d) ReaxFF

[Ref. 35], and (e) HyBOP [developed in this study]. The Wulff shape derived from DFT calculations are provided for reference in panel (f).

*4.4 Prediction of bulk gold properties*

In addition to describing clusters accurately, a good description of bulk properties is essential for a wide range of fundamental research problems including adsorption of Au nanoclusters on Au surfaces, diffusion of small Au clusters on surfaces of bulk Au or large clusters, and breakdown of large Au clusters into small ones during high energy impact. The inclusion of both SR and LR interactions enables our HyBOP to describe the condensed bulk phases of Au in excellent agreement with DFT calculations and experimental measurements. **Table 3** provides the structural, elastic, and cohesive energies of various bulk Au polymorphs predicted by our HyBOP with those obtained from original BOP parameters as well as DFT/experiments. Our HyBOP preserves the DFT evaluated energetic ordering of the various bulk polymorphs similar to the original BOP. In the framework of both the EFF models, FCC is the most stable bulk polymorph in agreement with previous DFT and experiments. Our HyBOP predicts lattice parameter of FCC Au to be 4.2 Å, in good agreement with our DFT calculations (4.17 Å), and previous experiments (4.07 Å).[69] In addition, the HyBOP predicted cohesive energy for FCC Au (-3.82 eV/atom) matches remarkably well with experimental reports (-3.81 eV/atom).[47] However, DFT-PBE significantly underestimates this value (-2.97 eV) owing to its inadequate treatment of the dispersion effects in Au.[7, 46] Despite this issue, DFT-PBE correctly identifies the energetic order of bulk phases and is known to provide an accurate description of the surfaces and nanoclusters.[35, 46] Furthermore, the errors associated with our HyBOP predicted elastic constants are similar to those for original BOP.

**Table 3** Structural, energetic, and elastic properties of bulk polymorphs of gold as predicted by the HyBOP parameters obtained in this study. These predictions are compared with values from original BOP parameters (Ref. 68), our DFT calculations, and previous experiments (if available). $E_c^{fcc}$ refers to cohesive energy of FCC, $a^j$ to lattice parameter of cubic polymorph $j$, and $\Delta E_c^{j-fcc}$ is the difference of cohesive energy between polymorph $j$ and FCC. The quantities $C_{ij}$ are the values of elastic stiffness constants.

| | HyBOP (This study) | BOP (Ref. 68) | DFT (This study) | Experiment |
|---|---|---|---|---|
| $E_c^{fcc}$ (eV/atom) | -3.82 | -3.81 | -2.97 | -3.81[a] |
| $a^{fcc}$ (Å) | 4.19 | 4.07 | 4.17 | 4.07[a] |
| $\Delta E_c^{bcc-fcc}$ (eV/atom) | 0.08 | 0.04 | 0.02 | - |
| $a^{bcc}$ (Å) | 3.34 | 3.23 | 3.31 | - |
| $\Delta E_c^{sc-fcc}$ (eV/atom) | 0.5 | 0.28 | 0.20 | - |
| $a^{sc}$ (Å) | 2.82 | 2.68 | 2.76 | - |
| $\Delta E_c^{dia-fcc}$ (eV/atom) | 1.30 | 1.0 | 0.71 | - |
| $a^{dia}$ (Å) | 6.26 | 5.98 | 6.18 | - |
| $C_{11}$ (GPa) | 231 | 201 | 150 | 192[b] |
| $C_{12}$ (GPa) | 170 | 151 | 129 | 163[b] |
| $C_{44}$ (GPa) | 75 | 47 | 31 | 42[b] |

[a]Ref. 23; [b]Ref. 47

## 5. CONCLUSIONS

In summary, we introduce a novel hybrid bond order potential (HyBOP) for Au, which accurately captures both the bond-directionality effects over short-distances, as well as long-range dispersion effects. The short-range interactions are described by Tersoff-type bond-order potential, while the long-range interactions are accounted via a scaled pairwise Lennard-Jones term whose contribution is dictated by the local atomic density. Such a hybrid bond order potential allows for an accurate description of Au systems across different length scales ranging from small clusters to surfaces and bulk condensed phases. To obtain optimum values of the independent parameters of our HyBOP, we outline a force-field fitting strategy composed of a combination of genetic algorithms and local optimization using simplex. Our newly developed HyBOP accurately describes (a) structure, dimensionality, energetics, and dynamics of Au nanoclusters, (b) critical cluster size for transition from planar to globular configurations, (c) surface energetics and consequently Wulff shapes for large clusters, and (d) structure, energetics, and elastic properties of bulk polymorphs of Au. Using this newly developed HyBOP, we performed extensive searches for global energy-minimum configurations at several sizes ($n \leq 20$) for few-atom Au clusters using evolutionary methods for structural optimization. From these searches, we found that our HyBOP predicts global minimum configurations (at all cluster sizes) consistent with previous DFT calculations and spectroscopic measurements. More importantly, these parameters describe size-dependent evolution of structural motifs (e.g., planar, hollow cages, space-filled structures) in $n < 20$- atom Au clusters, in perfect accordance with previous DFT studies. Furthermore, the inclusion of long-range interactions enables our HyBOP to describe the energetics/atomic-scale dynamics of surfaces and bulk condensed phases. Finally, our force-field fitting strategy and the HyBOP developed in this work will be a valuable tool for investigating atomic scale processes in Au based nano-systems for various energy and device applications.

## AUTHOR INFORMATION

**Corresponding Author**

*Email skrssank@anl.gov, mchan@anl.gov

## ACKNOWLEDGMENT

We acknowledge C. Wolverton and J. Greeley for helpful discussion regarding the use of genetic algorithm for force field fitting. Use of the Center for Nanoscale Materials was supported by the U. S. Department of Energy, Office of Science, Office of Basic Energy Sciences, under Contract No. DE-AC02–06CH11357. This research used resources of the National Energy Research Scientific Computing Center, a DOE Office of Science User Facility supported by the Office of Science of the U.S. Department of Energy under Contract No. DE-AC02-05CH11231. We gratefully acknowledge the computing resources provided on Blues and Fusion, a high-performance computing cluster operated by the Laboratory Computing Resource Center at Argonne National Laboratory.